%% file: main.tex
\documentclass[conference,a4paper]{IEEEtran}
\IEEEoverridecommandlockouts
\usepackage[hidelinks]{hyperref}
\usepackage[cmex10]{amsmath}
\usepackage{amssymb,amsfonts}
\usepackage{dblfloatfix}

\usepackage[ruled,vlined]{algorithm2e}
\usepackage{graphicx}
\graphicspath{{Figures/PDF/}{Figures/PNG/}}

\usepackage{booktabs}
\usepackage{siunitx}
\usepackage[numbers,compress]{natbib}
\usepackage{texnames}
\usepackage{bm,bbm}
\usepackage{orcidlink}
\usepackage{subcaption}
\usepackage{multirow}

\begin{document}
\title{\uppercase{Control Copy-Paste: Controllable Diffusion-Based Augmentation Method for Remote Sensing Few-Shot Object Detection}}

\author{
\IEEEauthorblockN{Yanxing Liu\orcidlink{0009-0007-8604-933X}}
	\IEEEauthorblockA{
		\textit{Key Laboratory of Technology in Geo-spatial} \\ 
		\textit{InformationProcessing and Application System,}\\
		\textit{Aerospace Information Research Institute,}\\
		\textit{Chinese Academy of Sciences} \\
		100190 Beijing, China\\
		liuyanxing21@mails.ucas.ac.cn}
	\and
	\IEEEauthorblockN{Jiancheng Pan\orcidlink{0000-0001-5968-5209}}
	\IEEEauthorblockA{
		\textit{Department of } \\
		\textit{Earth System Science,} \\
		\textit{Tsinghua University}\\
		100084 Beijing, China\\
		jiancheng.pan.plus@gmail.com}
	\and
	\IEEEauthorblockN{Bingchen Zhang}
	\IEEEauthorblockA{
		\textit{Key Laboratory of Technology in Geo-spatial} \\ 
		\textit{InformationProcessing and Application System,}\\
		\textit{Aerospace Information Research Institute,}\\
		\textit{Chinese Academy of Sciences} \\
		100190 Beijing, China\\
		zhangbc@aircas.ac.cn}
}
\vspace{-15mm}

\maketitle
\begin{abstract}
Few-shot object detection~(FSOD) for optical remote sensing images aims to detect rare objects with only a few annotated bounding boxes. 
The limited training data makes it difficult to represent the data distribution of realistic remote sensing scenes, which results in the notorious overfitting problem. 
Current researchers have begun to enhance the diversity of few-shot novel instances by leveraging diffusion models to solve the overfitting problem.
However, naively increasing the diversity of objects is insufficient, as surrounding contexts also play a crucial role in object detection, and in cases where the object diversity is sufficient, the detector tends to overfit to monotonous contexts. 
Accordingly, we propose Control Copy-Paste, a controllable diffusion-based method to enhance the performance of FSOD by leveraging diverse contextual information. 
Specifically, we seamlessly inject a few-shot novel objects into images with diverse contexts by a conditional diffusion model.
We also develop an orientation alignment strategy to mitigate the integration distortion caused by varying aspect ratios of instances. 
Experiments on the public DIOR dataset demonstrate that our method can improve detection performance by an average of 10.76\%.
\end{abstract}


\input{tex/introduction.tex}

\input{tex/method.tex}

\input{tex/experiment.tex}

\input{tex/conclusion.tex}

\bibliographystyle{IEEEtranN}
\bibliography{references}

\end{document}

%% file: tex/introduction.tex
\section{Introduction}
\label{sec:intro}
During the past decade, object detection has made significant progress, thanks to the remarkable progress of deep artificial neural networks. 
However, the acquisition of remote sensing images~(RSIs) is limited compared to natural scene images, as remote sensing images are often captured by satellites or drones.
Moreover, object-level annotation is more challenging owing to the RSIs' high resolution and complex background, which exhibits more information redundancy~\citep{pan2023reducing,10507076}in the semantic space~\cite{pan2023prior,pan2024pir}. 
As a result, collecting large-scale labelled data for training is time-consuming or even unfeasible for many applications.
If the few-shot training data is directly applied for training, detectors will suffer from the notorious overfitting problem.
To this end, few-shot object detection~(FSOD) has been proposed, aiming to achieve object detection tasks with a limited number of training samples. \par{}

FSOD seeks to leverage knowledge learned from data-rich base datasets to enhance detection performance for few-shot novel classes. Concretely, most FSOD approaches achieve it via transfer learning~\citep{wang2020frustratingly,sun2021fsce,pan2025enhancesearchaugmentationsearchstrategy,fu2025ntire} or meta-learning~\citep{kang2019few,wang2019meta}. 
Methods based on transfer learning transfer the knowledge from base datasets to few-shot novel datasets. 
Wang et al.~\citep{wang2020frustratingly} proposed a frustratingly simple few-shot object detection approach that fine-tunes only the last layer of existing detectors for rare classes. 
FSCE~\citep{sun2021fsce} further proposed a novel method to learn contrastive-aware object proposal encodings to classify detected objects. 
Kang et al.~\citep{kang2019few} designed a novel FSOD approach, which automatically reweights features of novel classes based on features of support images. 
Based on Faster R-CNN~\citep{ren2016faster}, Meta-RCNN~\citep{wang2019meta} proposed meta-learning over Region of Interest~(RoI) features and enabled detectors to acknowledge class-agnostic priors. 
LAE-DINO~\citep{pan2024locate} employs open-vocabulary detection to utilize prior knowledge in the text encoder.
While meta-learning and transfer learning methods are straightforward, these methods were still trained on limited training data, where the inherent overfitting problem remains unsolved. \par{}
\begin{figure}[bt]
	\centering
	\includegraphics[width=\linewidth]{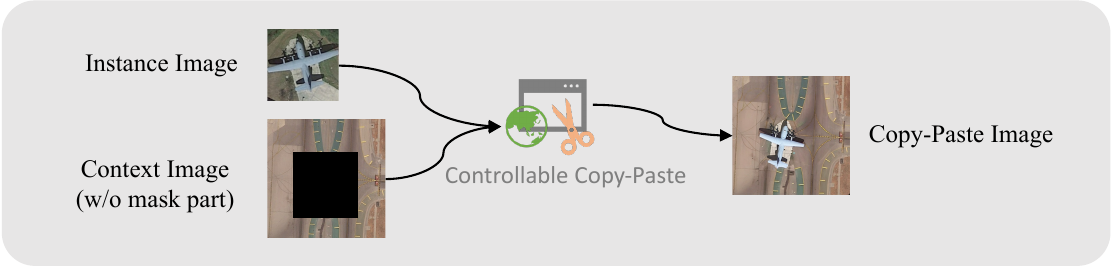}
	\caption{The illustration of the proposed Control Copy-Paste pipeline.}\label{fig:fig0} \vspace{-5mm}
\end{figure}
Recently, researchers~\citep{lin2023explore,zhang2024advancing,pan2025earthsynth} have begun to utilize existing text-to-image diffusion models~\citep{rombach2022high} to address the challenge of FSOD. 
They employ diffusion models to generate diverse instances and use them to train detectors. 
However, they overlook the significance of context in the FSOD task. 
In our experiments, we observe that detectors tend to overfit complex contexts in the FSOD task of RSIs, and the harmonious transfer of few-shot instances into diverse contexts can significantly improve the performance of FSOD in RSIs.
Therefore, based on~\citep{chen2024anydoor}, we propose a pipeline to improve the detection performance by harmoniously improving the diversity of contexts, as illustrated in Fig.~\ref{fig:fig0}. 
Compared with traditional data-augmentation methods like copy-paste~\citep{ghiasi2021simple}, our method can inject few-shot novel instances into the context more harmoniously, which enables the detector to utilize the features of the context better.
Our approach first trains a class-agnostic instance integration network on base datasets to enable the model to extract common features of RSIs and then fine-tunes it on a few-shot novel instances to improve the integration performance further.
Through comprehensive experiments on the DIOR dataset, our method improves the detection performance of few-shot classes by an average of 10.76\%. \par{}
In conclusion, the contributions of this paper are summarized as follows.
\begin{enumerate}
	\item Based on extensive experiments, we find that when training data is limited, detection performance is constrained by both context and object diversity. In RSIs, large-scale context pixels increase the susceptibility to context overfitting.
	\item We propose the Control Copy-Paste, a controllable, diffusion-based pipeline to enhance contextual diversity by seamlessly integrating objects into diverse contexts via diffusion models.
	\item By integrating our pipeline with different FSOD approaches, experimental results show that seamlessly integrating the target into diverse contexts leads to significant performance improvements in FSOD for RSIs.
\end{enumerate}

%% file: tex/method.tex
\section{Methodology}
\label{sec:metho}

\subsection{Preliminary}
In FSOD, detectors are first trained using base classes with sufficient labelled images to learn class-agnostic features and subsequently trained to detect $N$ novel classes ($N$-way) using only $K$ annotated instances ($K$-shot). 
The whole dataset consists of two types of classes: base classes $C_b$ and novel classes $C_n$, where $C_b \cap C_n = \emptyset$. \par{}

\subsection{What limits the performance of FSOD: An Analysis of Factors Influencing FSOD}
\label{sec:metho.2}
\begin{figure}[bt]
	\centering
	\includegraphics[width=\linewidth]{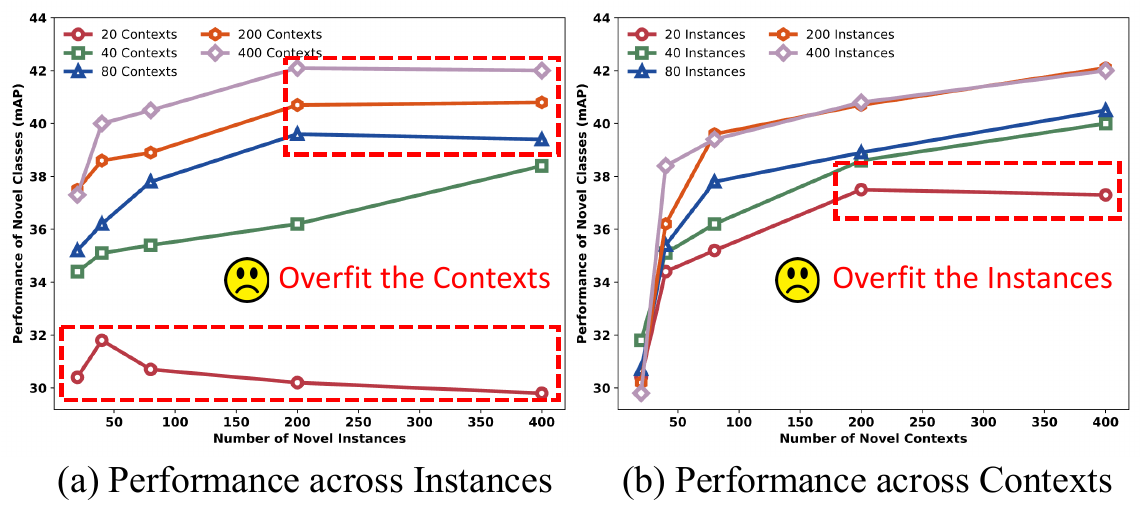}
	\caption{Analysis of different components. (a). Detection performance varies across different instances. (b). Detection performance varies across different contexts.}\label{fig:fig1}
\end{figure}

To further improve the performance of FSOD, we have investigated the impact of both object context and object itself on the FSOD performance of RSIs. 
We adopt the method of copy-paste~\citep{ghiasi2021simple} to decouple the contexts and instances. 
The key idea is to inject different object instances into different context images by copy-paste. 
By separately fixing the number of instances and contexts, and training the enhanced data using FSCE~\citep{sun2021fsce}, we get results in Fig.~\hyperref[fig:fig1]{2(a)} and Fig.~\hyperref[fig:fig1]{2(b)}. \par{}
As demonstrated in Fig.~\hyperref[fig:fig1]{2(a)}, we observe that when the object contexts are limited (20-shot per class), simply increasing the number of instances fails to improve the FSOD performance. 
When the number of instance contexts increases from 20 to 40, the FSOD performance of the detector demonstrates an improvement.
This indicates that the lack of contextual diversity can also limit the performance of FSOD.
In Fig.~\hyperref[fig:fig1]{2(b)}, the performance of the detector improves with the increase of contexts, further demonstrating that enhanced context diversity can prevent the model from overfitting. 
We also observe the performance saturation in both Fig.~\hyperref[fig:fig1]{2(a)} and Fig.~\hyperref[fig:fig1]{2(b)}, where the performance nearly remains constant despite increased diversity and we argue that the reason for this is that the model has overfitted to another element.
For instance, when the diversity of contexts is sufficient but the diversity of objects is insufficient, the model tends to overfit the objects themselves, and vice versa. 
Furthermore, our analysis reveals that in remote sensing scenarios, models are more prone to overfitting to contexts owing to the larger contextual scenes, as evidenced by the earlier overfitting of contexts.\par{} 

\begin{figure*}[bt]
	\centering
	\includegraphics[width=0.8\linewidth]{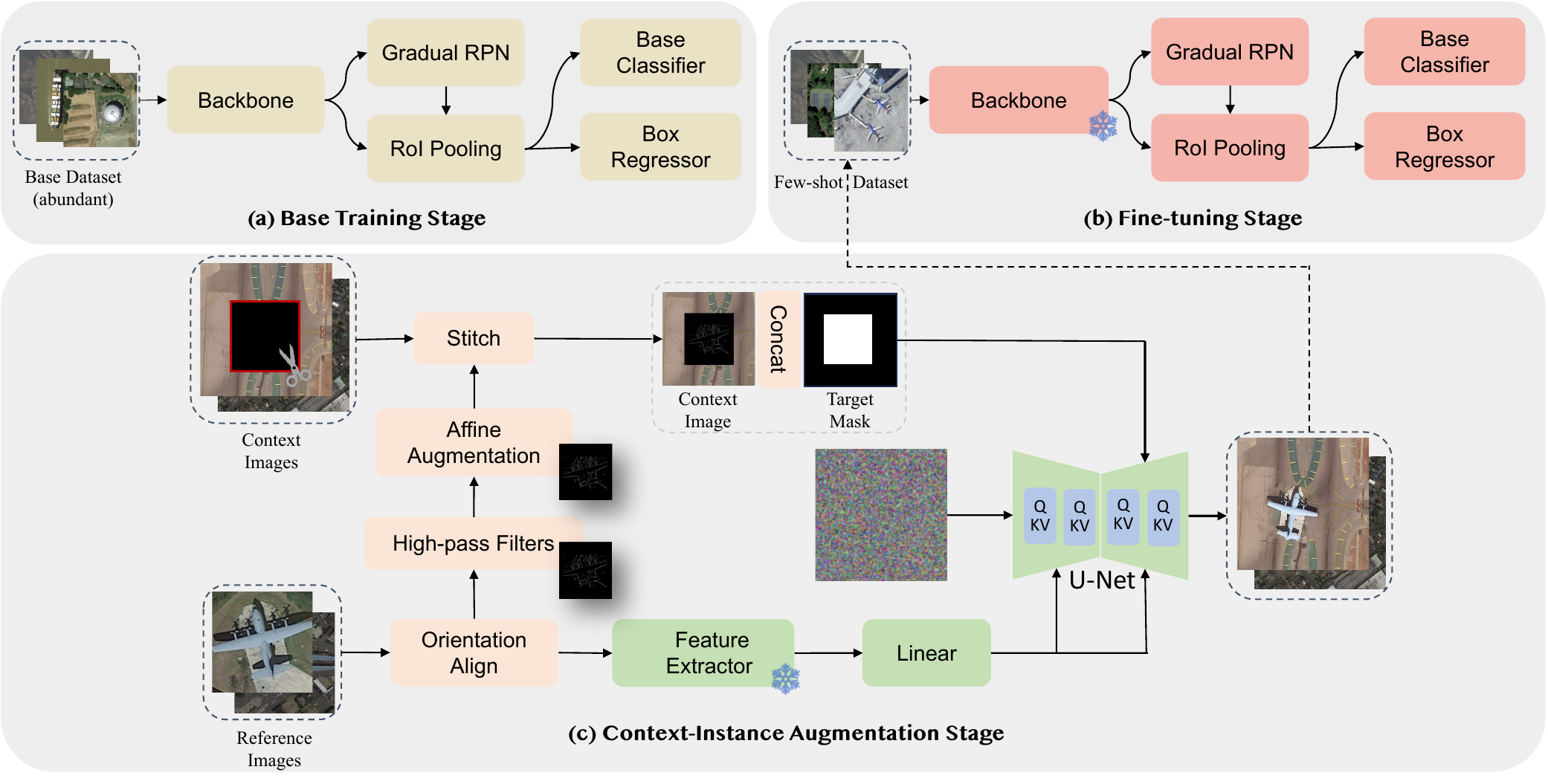}
	\caption{The architecture framework of the proposed pipeline for few-shot object detection on remote sensing images. The detector is first trained on a base dataset containing abundant objects to learn domain-agnostic knowledge of RSIs. Few-shot novel objects are then injected into multiple context images through a class-agnostic conditional diffusion model. Finally, the synthetic dataset is combined with the few-shot dataset to fine-tune the detectors.}\label{fig:fig2}
	\vspace{-2mm}
\end{figure*}

\begin{table*}[hbt]
	\centering
	\caption{Comparison of 5-way FSOD performance on the novel classes of DIOR dataset. } \label{tab:tab1}
	
	\begin{tabular}{ccccccccc}
	\toprule
	Shots & Method & mAP(\%) & $\Delta$ & Airplane & Baseball field & Tennis court & Train station & Windmill \\ \hline
	\multirow{6}{*}{3-shot}	  & FSCE   & 18.30 & -  & 7.70     & 41.00   & 22.30 & 11.50 & 9.10 \\
		  & MSOCL & 25.75 & - & 17.38     & 47.15   & 42.98 & 8.68 & 12.56 \\ 
		  & FSCE+Copy-Paste & 28.31 & $\uparrow$10.01  & 18.50 & 45.60 & 59.10 & 6.60  & 11.60                     \\
		  & MSOCL+Copy-Paste & 36.61 & $\uparrow$10.86  & 27.87 & 53.65 & 64.77 &  \textbf{13.96} & 22.83  \\
		  & FSCE+Ours 		& 32.18 & $\uparrow$\textbf{13.88} & 21.30   & 46.80   & 58.50 & 7.90 & \textbf{26.30}          \\
		  & MSOCL+Ours & \textbf{38.79} & $\uparrow$13.04  & \textbf{35.99} & \textbf{56.85} & \textbf{67.39} & 10.38  & 23.36  \\ \hline
	\multirow{6}{*}{5-shot}	  & FSCE   & 18.59 & -  & 9.10 & 41.10 & 20.40 & 13.00 & 9.40 \\
			& MSOCL & 26.16 & -   & 15.95     & 49.29   & 34.03 & \textbf{17.79} & 13.73 \\
		  & FSCE+Copy-Paste & 30.87 & $\uparrow$12.28  & 16.20 & 48.50 & 59.30 & 12.00 & 18.30                    \\
		  & MSOCL+Copy-Paste & 38.36 & $\uparrow$12.10  & 31.15 & \textbf{57.84} & \textbf{66.44} &  14.59 & 21.76  \\
		  & FSCE+Ours 		& 33.21 & $\uparrow$\textbf{14.62}   & 21.20 & 50.00  & 57.70 & 12.20 & 25.00                    \\ 
		  & MSOCL+Ours & \textbf{39.99} & $\uparrow$13.83  & \textbf{37.45} & 57.59 & 65.42 & 12.71  & \textbf{26.79}  \\ \hline
	\multirow{6}{*}{10-shot}	  & FSCE   & 26.16 & -  & 10.40 & 44.50 & 41.70 & 19.30 & 14.90 \\
	& MSOCL & 32.65 & -  & 18.52  & 51.75   & 54.86 & \textbf{19.76} & 18.36  \\
		  & FSCE+Copy-Paste & 33.20 & $\uparrow$7.04  & 17.10 & 48.80 & 58.70 & 17.00 & 24.60                    \\
		  & MSOCL+Copy-Paste & 39.45 & $\uparrow$6.80 & 31.26 & 57.30 & 65.14 &  16.25 & 27.30  \\
		  & FSCE+Ours 		&  34.52 & $\uparrow$8.36  &  21.40   &  50.90  & 53.80 & 17.40 & 29.10                    \\ 
		  & MSOCL+Ours & \textbf{41.70} & $\uparrow$\textbf{9.14}  & \textbf{37.90} & \textbf{57.77} & \textbf{66.07} & 16.13  & \textbf{30.64}  \\ \hline
	\multirow{6}{*}{20-shot}	  & FSCE   & 29.60 & -  & 15.60 & 45.50 & 48.20 & 22.00 & 16.70  \\
	& MSOCL & 37.36 & -   & 30.26   & 52.33   & 57.68 & 21.14 & 25.39 \\
		  & FSCE+Copy-Paste & 34.50 & $\uparrow$4.90  & 19.50 & 47.30 & 57.90 & \textbf{23.30} & 24.80                     \\
		  & MSOCL+Copy-Paste &  40.68 & $\uparrow$3.32 & 33.87 & 57.97 & 65.28 & 18.12  & 28.19  \\
		  & FSCE+Ours 		& 35.78 & $\uparrow$6.18   & 20.80  & 49.00   & 57.10 & 22.90 & 29.20                     \\ 
		  & MSOCL+Ours & \textbf{44.46} & $\uparrow$\textbf{7.10}  & \textbf{41.85} & \textbf{58.82} & \textbf{65.43} & 17.93  & \textbf{38.27}  \\ \hline
	\bottomrule
	
	\end{tabular}

\end{table*}

\subsection{Control Copy-Paste Pipeline}
As analyzed in section~\ref{sec:metho.2}, both context diversity and instance diversity contribute to the high detection performance of the detector based on deep learning. 
Insufficient diversity in any single element can lead to model overfitting, and the detector is more likely to overfit the context of RSIs owing to the larger number of pixels in the context image. 
To this end, based on AnyDoor~\citep{chen2024anydoor}, we propose a pipeline to harmoniously integrate few-shot novel instances into diverse contexts to avoid overfitting to few-shot contexts. \par{}
We adopt the two-stage training strategy that is widely adopted in FSOD. 
As demonstrated in stage I of Fig.~\ref{fig:fig2}(a), the detector is trained on instances belonging to base classes to extract common features of RSIs. 
Then, as shown in Fig.~\ref{fig:fig2}(b), the detector is trained on few-shot novel instances to achieve the detection for novel classes. 
We adopt the Gradual RPN\cite{liu2024few} to enhance the locating accuracy of the model.
In addition, the synthetic data generated by our pipeline is also incorporated into the training data to provide diverse contexts for few-shot novel instances. 
Note that the pipeline is approach-agnostic and can be applied to various FSOD approaches. \par{}
The context augmentation pipeline is demonstrated in Fig.~\ref{fig:fig2}(c). Given the few-shot reference image, the context image, and the location where the reference image should be integrated, the pipeline will produce a synthetic image for training.
The pipeline is flexible, as both reference images and context images can be derived from real scenes or other T2I models~\citep{rombach2022high}.
The features of the reference image are characterized by both coarse-grained and fine-grained features. 
The coarse-grained features are obtained from a robust self-supervised DINOv2~\citep{oquab2023dinov2}, while the fine-grained features are derived from high-pass filtering. 
The reference image is resized to $I_r  \in \mathbb{R}^{224 \times 224}$ at first. To maintain the aspect ratio of the reference instance, the longer side is directly resized to 224 pixels while the shorter side is padded to match the length of the longer side before being resized to 224 pixels.
The coarse feature extractor encodes $I_r$ to obtain the coarse feature, which is in $\mathbb{R}^{257 \times 1536}$, which is then mapped into the embedding space of U-Net through a linear layer.
Due to the lack of spatial detail in coarse-grained features, the pipeline utilizes high-frequency edges as detailed guidance for the target instance.
Given the diversity of imaging angles, instances in RSIs often appear from multiple perspectives, and we implement affine transformations, including rotation, mirroring, and flipping, to enhance the diversity of reference instances.
The transformation of the reference image can be formulated as
\begin{align}
	\label{eq:eq0}
	I_s = F_{Aff} \circ F_{Hpf} \circ F_{O}(I_r),
\end{align}
where $I_s$ is the image to be stitched, $F_{Aff}()$, $F_{Hpf}$, and $F_{O}$ denote the affine transformation, high-pass filtering, and orientation alignment operation, respectively.
Finally, the coarse-grained features of the reference instance and the context image containing the fine-grained features of the target serve as conditions for stable diffusion~\citep{rombach2022high}. The reverse diffusion process is formulated as 
\begin{align}
	\label{eq:eq0.1}
	p(z_{t-1} | z_t, x_c, m_c, x_r) = UNet(z_t, x_c \odot m_c, x_r),
\end{align}
where $t$ is the timestep in diffusion, $x_c$ and $m_c$ are features and mask of context image, $x_r$ is features of reference image.
\par{}
Since the aspect ratios of many remote sensing objects are quite extreme, aligning the orientation of the reference image with the target is crucial for achieving satisfactory performance. 
Thus, we align the long edges of reference images and the target areas to ensure the consistent orientation between reference images and target areas. 
\begin{align}
	\label{eq:eq1}
	I_r^{\prime} = \begin{cases}
		Rotate(I_r, \frac{\pi}{2}), & (R_r-1)(R_t-1)<0 \\
		I_r, & else 
	\end{cases}
\end{align}
The detailed calculation is formalized in Eq.~\ref{eq:eq1}, where $I_r$ denotes the reference images, $R_r$ and $R_t$ represent the aspect ratios of the reference instances and target area, respectively.
 \par{}
The pipeline is class-agnostic, so we can apply it to the base dataset to enhance the feature representation of objects.
After base training, the model undergoes few-shot fine-tuning for novel instances to improve the integration performance further.

%% file: tex/experiment.tex
\section{Results}
\label{sec:res}
\subsection{Datasets and Implementation details}
We perform comprehensive experiments on the DIOR~\citep{li2020object} dataset. 
The dataset contains 20 classes divided into 15 base classes and 5 novel classes. 
In experiments, the airplane, tennis court, train station, baseball field, and windmill are set as novel classes, and the remaining classes are base classes. 
K-shot (K = 3, 5, 10, 20) instances of novel classes are sampled as training data to simulate a realistic scene with few shots. 
The mean average precision (mAP@0.5) introduced in~\citep{everingham2010pascal} is used for evaluating the performance of a single class. 
We perform training on the trainval set of the DIOR dataset and evaluate the performance of novel classes on the test set. 
The context images are sourced from the DIOR dataset, where we have verified that all scene images in the dataset do not contain any novel instances. 
To avoid hallucination, the training data is combined with the few-shot dataset and the synthetic dataset. \par{}
\subsection{Result Analysis}
In this section, one classical natural scene FSOD method FSCE~\citep{sun2021fsce} and one FSOD approach for RSIs, named MSOCL~\citep{9984671} are adopted. 
The experiment results have been reported in Table~\ref{tab:tab1}. 
As the table shows, both copy-paste and our proposed method can improve the performance of novel classes by increasing context diversity. 
With only three instances and forty contexts, the detector demonstrates performance comparable to that of the detector using the $20$-shot dataset. 
This indicates that contextual information plays a crucial role in object detection for RSIs, and increasing contextual diversity can mitigate the risk of overfitting specific contextual patterns. 
Although copy-paste can enhance contextual diversity, the box-level annotation constrains the ability to fully decouple context and target instances, thus limiting its performance. 
By completely decoupling contexts and instances, our method can further achieve up to 3.81\% performance improvement over Copy-Paste.
Copy-paste and our proposed method both lead to a slight performance drop in the train station class. 
Train stations have extremely high aspect ratios, and misalignment between the reference image and the context area can cause a performance drop.
Although we have mitigated this to some extent through the orientation alignment strategy, the performance of the train station remains unsatisfactory. 
However, our method still achieves an average performance improvement of 10.76\% by enhancing the performance of most classes.

%% file: tex/conclusion.tex
\section{Discussion}
\label{sec:dis}
In this work, we explore the critical role of context in FSOD for RSIs and propose a data augmentation method leveraging diffusion models.
We find that simply increasing the diversity of instances is not sufficient to mitigate overfitting, and even with sufficient objects, detectors still become overfitted to the monotonous context. 
Although copy-paste can also increase context diversity, the use of box-level annotation makes it impossible to fully decouple the context from the instance itself, thereby limiting performance.
Inspired by~\citep{rombach2022high,chen2024anydoor}, we propose a data augmentation method based on diffusion models, which enhances the performance of FSOD in RSIs by harmoniously injecting the target objects from the few-shot training set into different contexts.
The experimental results indicate that our pipeline outperforms copy-paste in various FSOD approaches. \par{}
Although increasing contextual diversity can enhance the performance of FSOD in RSIs, we have found that this has limitations. 
When the context diversity is sufficient, the diversity of the object itself becomes the primary factor limiting the performance of detectors. 
This suggests that increasing the diversity of the object itself is also important for improving detection performance.
We hope our work will gain a deeper understanding of the overfitting phenomenon in FSOD of RSIs and further enhance the performance of detectors in the scenario of few training data.